\documentstyle[12pt]{article}
\begin{document}
\def\gd{\delta}
\def\ga{\alpha}
\def\gb{\beta}
\def\gg{\gamma}
\def\go{\omega}
\def\dya{\frac{\partial}{\partial y^\ga }}
\def\dyb{\frac{\partial}{\partial y^\gb }}
\def\dzi{\frac{\partial}{\partial z^i }}
\def\dzj{\frac{\partial}{\partial z^j }}
\def\ge{\varepsilon}
\def\gk{\kappa}
\def\half{\frac{1}{2}}

\vspace{-1mm}
\begin{flushright} G\"{o}teborg ITP 93-42 \\
\vspace{-1mm}
 FIAN/TD/18--93\\
\vspace{-1mm}
{September 1993}\end{flushright}\vspace{2cm}

\begin{center}
{\Large{\bf Representations of the \\
$S_N$-Extended Heisenberg Algebra and Relations Between \\
Knizhnik-Zamolodchikov Equations \\
 and Quantum Calogero Model }}\vspace{15mm}\\
{\bf\large Lars ~Brink }\vspace{5mm}\\
Institute for Theoretical Physics, S-412 96 G\"{o}teborg,
Sweden\vspace{1.5cm}\\
{\bf\large Mikhail A.~Vasiliev$\,
{}^\dagger
$}
\vspace{0.5cm}\\
I.E.~Tamm Theoretical Department, P.N.~Lebedev
Physical Institute, \\117924, Leninsky Prospect 53, Moscow, Russia
\vspace{2cm}\\
{\bf Abstract}
\end{center}
\noindent
We discuss lowest-weight representations of the $S_N$-Extended
Heisenberg Algebras underlying
the $N$-body quantum-mechanical Calogero model. Our construction leads
to flat derivatives interpolating between Knizhnik-Zamolodchikov
and Dunkl derivatives. It is argued that based on these results one
can establish new links between solutions of the Knizhnik-Zamolodchikov
equations and wave functions of the Calogero model.

\vspace{3cm}
\noindent
---------------------------------

${}^\dagger$
{\footnotesize Supported in part by the Russian Fund of Fundamental Research,
grant N67123016.}

\newpage

Recently, it was shown \cite{BHV} that the N-body quantum-mechanical Calogero
model with a harmonic potential can be interpreted as a free model of
oscillators $a_i ,a^+_j \quad (i,j = 1,...,N)$ obeying the following
commutation
relations
$$ [a^{(+)}_i ,a^{(+)}_j ]=0 \,,\quad \,
[a_i ,a^{+}_j ]= \gd_{ij }(1+\nu\sum_{l}K_{il})-\nu K_{ij}\,, \eqno{(1)}
$$
where $\nu$ is a constant related to the Calogero coupling constant, while
            $K_{ij}$    are operators obeying the exchange algebra
$$
K_{ij}K_{jl}=K_{jl}K_{il}=K_{il}K_{ij} \eqno{(2)}
$$
for all $i\neq j,\,i\neq l,\,j\neq l$  (no summation over repeated indices), $$
(K_{ij})^2=I\,,\qquad K_{ij}=K_{ji}\,, \eqno{(3)}
$$ and the following relations are true
$$ K_{ij}a^{(+)}_j=a^{(+)}_i K_{ij}.  \eqno{(4)}
$$

The algebra (1)-(4), which can be called the
$S_N$-extended
Heisenberg algebra
will be denoted as $SH_N$. It
admits the following
realization \cite{POL1,BHV} $$
a^\mp_i = \frac{1}{\sqrt{2}} (z_i \pm D_i) \eqno{(5)}
$$ with
$$ D_i =\frac{\partial }{\partial z_i}+\nu \sum_{j\neq i}(z_i -z_j)^{-1}
(1-K_{ij})\,, \eqno{(6)}
$$ which enables one to solve the Calogero model by observing \cite{BHV} that
for the subsector  of identical particles the Calogero Hamiltonian can be cast
into the form $$
H = \half \sum_{i} \{a_i \,,a^+_i\} \eqno{(7)} $$ and
$$ [H, a^+_i\,] = a^+_i\,,\qquad [H, a_i\,] = -a_i \,\,.\eqno{(8)}
$$

The remarkable property of the derivatives $D_i$ is that they are mutually
commuting
$$
[D_i , D_j ] = 0
\eqno{(9)}
$$
as was first shown in the mathematical literature (see \cite{DUN} and
references
therein) and later rediscovered in \cite{POL1,BHV} in relation with the
Calogero problem.  In fact, $D_i$ obey the same algebra with $z_i$ as
$a^{(+)}_i
$ do
$$ [D_i ,z_j ]= \gd_{ij }(1+\nu\sum_{l}K_{il})-\nu K_{ij}
\eqno{(10)}
$$
$$ K_{ij} D_j=D_i K_{ij}\,,\qquad K_{ij} z_j=z_i K_{ij}\,,
\eqno{(11)}
$$
because the right hand sides of the formulae (1) and (10) are
symmetric.

The Dunkl derivatives (6) are also very important for the Calogero model
without oscillator interaction (i.e. that with continuous spectrum) since
they give rise to the commuting conserved currents of the model \cite{POL1}.

The relations (5), (6) give a particular representation of the algebra
(1)-(4). The question we address in this letter is whether there exist
other representations of the same algebra, inequivalent to the
representation (5), (6). We show that such representations do actually
exist and  can be useful in the context of the
interplay between the Calogero problem and Knizhnik-Zamolodchikov equations
\cite{KZ}.

The basic element of our construction is a straightforward generalization of
the standard construction of the Fock space for the ordinary Heisenberg
algebra. Namely, let us consider lowest weight-type representations admitting
a vacuum subspace $|0\rangle$ obeying the conditions
$$
a_i |0\rangle = 0\,,
\eqno{(12)}
$$
which form $m\times m$ irreducible representations of the zero-grade
subalgebra, the group algebra of $S_N$ spanned by all combinations of products
of $K_{ij}$.  This means that $|0\rangle$ is the $m$-column such that $$
K_{ij} |0\rangle = T_{ij} |0\rangle\,,
\eqno{(13)} $$
where $T_{ij}$ is the $m\times m$ matrix representation of the elementary
$i\leftrightarrow j$ permutation operators of $S_N$. Consistency of eq.(13)
implies that the following relations have to be true:
$$
(T_{ij})^2=I\,,\qquad T_{ij}=T_{ji}
\eqno{(14)}
$$
$$
T_{ij}T_{jl}=T_{jl}T_{il}=T_{il}T_{ij}
\eqno{(15)}
$$
$$
K_{ij}T_{jl}=T_{il}K_{ij}\,,\, \qquad T_{ij} K_{ij} =K_{ij}T_{ij}
\eqno{(16)}
$$
for all $i\neq j,\,i\neq l,\,j\neq l$  (no summation over repeated indices).
The whole representation is then given by the Fock-type module
$$
(a^+_1 )^{n_1}\ldots (a^+_N )^{n_N} |0\rangle\,.
\eqno{(17)}
$$

Sometimes it is more convenient to apply such a procedure
to the derivatives $D_i$ and coordinates $z_i$
themselves instead of the
creation and annihilation operators
requiring
$$
D_i |0\rangle = 0
\eqno{(18)}
$$
in place of (12). Algebraically, there is no difference since the operators
$z_i$ and $D_j$ obey the same algebra (9),(10) as $a_i$ and $a^+_j$ do. In
practice this directly leads to the formula (6) provided that one identifies
$|0\rangle$ with the constant 1, and the module
$(z_1 )^{n_1}\ldots (z_N )^{n_N} |0\rangle$ with the basis monomials of the of
the
linear space of polynomial functions. Proceeding along the same lines for the
general vacuum conditions (13), (18) one arrives at the following
generalization of (6)
$$
D_i =\frac{\partial }{\partial z_i}+\nu \sum_{j\neq i}(z_i -z_j)^{-1}
(T_{ij}-K_{ij})\,.
\eqno{(19)}
$$
Using (2), (3), the second condition in (11) and (14)-(16), it is not
difficult to check that the operators $D_i$ along with $z_i$ and $K_{ij}$
do indeed form the algebra (9)-(11) \footnote{To avoid misunderstandings let us
stress once more that $T_{ij}$ are $m\times m$ constant matrices
$(T_{ij})_\ga^\gb$ ($\ga \,, \gb = 1\ldots m$) commuting with the coordinates
$z_i$ and acting on the vacuum basis vectors $|0\rangle_\ga$.} and obey the
vacuum
condition (18). In particular, the $D_i\,$'s are still commuting.

The formula (19) which gives a new class of representations of the algebra
$SH_N$
constitutes the central result of this paper.

The following comments are now in order:

1. One can define the operators
$$
k_{ij} = T_{ij} K_{ij} =K_{ij}T_{ij}
\eqno{(20)}
$$
such that
$
[k_{ij} , T_{nm}] =0
$
as a consequence of (15), (16). Then, the operator (19) takes the form
$$
D_i =\frac{\partial }{\partial z_i}+\nu \sum_{i\neq j}(z_i -z_j)^{-1}
T_{ij}(1-k_{ij})\,.
\eqno{(21)}
$$
{}From (11), (19) and the fact that $T_{ij}$ commutes with $z_i$ it
follows that $k_{ij}$ acts as the operator interchanging the
coordinates $z_i \leftrightarrow z_j$ proportional to the unit matrix
in the $m$-dimensional vector space, $k_{ij} |0\rangle=|0\rangle$. This result
is quite natural since it is the combination $(z_i -z_j )^{-1} (1-k_{ij})$
which
leaves the space of polynomials $P(z)$ (i.e. the Fock space) invariant
since $(z_i -z_j )^{-1}(P(\ldots z_i \ldots z_j \ldots)-P(\ldots z_j \ldots z_i
 \ldots))$ is once again some polynomial. On the other hand, the operator
$k_{ij}$
can hardly be used itself in the practical calculations because it does not
rotate properly the derivatives (21) ($k_{ij} D_j \neq D_i k_{ij}$) and
therefore has no invariant meaning without referring to the specific
representation.

2. So far we required the representation of the basic algebra to be of the
lowest weight type, i.e. to obey (18) for some vacuum subspace $|0\rangle$. It
is this requirement that leads to the factor of $(1-k_{ij})$ on the
right-hand-side
of (21). If not insisting on this property, one can easily generalize
formula (19) to the following one

$$
D_i =\frac{\partial }{\partial z_i}+ \sum_{j\neq i}(z_i -z_j)^{-1}
(\mu T_{ij}-\nu K_{ij})\,.
\eqno{(22)}
$$

It is easy to check that these operators give some representation of
the same basic algebra (9)-(11) for arbitrary constants $\mu$ and $\nu$
provided that the relations (14)-(16) are true. For $\mu \neq \nu$ however
application of the $D_i$ to constant functions leads to proliferating
singularities.

For the case $\nu =0$ formula (22) can be interpreted as the
Knizhnik-Zamolodchikov operators in the case of unitary groups.
In these operators the exchange operators are
$T_{ij}= \sigma^a_i \sigma_{j\,a}
+ c$
where $\sigma^a_i$ is the generator for the unitary group acting
on a conformal field $\Phi (z_i )$, and $c$ is some constant. It is well
known that
$T_{ij}$ defined in this way
satisfies (14) and (15) for an appropriate constant $c$,
which, in turn,  can be removed by a
simple similarity transformation. Hence formula (22) interpolates between the
Knizhnik-Zamolodchikov operators $(\nu =0)$ and those underlying the Calogero
problem $\mu = \nu $ \cite{BHV} or $\mu = 0$ \cite{POL1}. One can expect that
these operators are indeed important  for the analysis of both of these
problems establishing links between Calogero-type integrable systems and
Knizhnik-Zamolodchikov equations (and, therefore, conformal models).

To illustrate this issue, let us consider the Hamiltonian of the form (7)
constructed from the derivatives (19) using (5). The explicit form of $H$
is
$$
H=\half \sum_{i=1}^N [-(D_i)^2 + (z_i)^2 ] =\nonumber
$$
$$
-\half [\Delta
+\nu \sum_{i\neq j} [(z_i -z_j)^{-1} (\dzi -\dzj) T_{ij}
- (z_i -z_j)^{-2} (T_{ij}-K_{ij})] -\sum_i z_i^2]\,.
\eqno{(23)}
$$
Eigenfunctions of the Hamiltonian (23) are given by the Fock vectors (17)
constructed through the derivatives (19). The vacuum wave function obeying
(12) has the standard form $ |0\rangle =(2\pi)^{-\frac{N}{2}}\exp(-\half \sum_i
z_i^2)$, so that it is a purely algebraic procedure to derive explicit
expressions for the eigenfunctions (17) as functions of $z$.

Now consider some matrix valued function $g(z)$ which is a solution of the
Knizhnik-Zamolodchikov equation
$$
\dzi g_\ga^\gb (z)= \nu g_\ga^\gg (z) \sum_{j\neq i} (z_i -z_j)^{-1}
(T_{ij})_\gg^\gb \,.
\eqno{(24)} $$
It is straightforward to check that $$
g^{-1} \Delta g = \Delta + 2\nu \sum_{i\neq j} (z_i - z_j)^{-1} T_{ij}\dzi -
\nu
\sum_{i\neq j} (z_i - z_j)^{-2} T_{ij} +\nu^2 \sum_{i\neq j} (z_i - z_j)^{-2}
\eqno{(25)} $$
and therefore $$
H= -\half [g^{-1} \Delta g  + \nu \sum_{i\neq j} (z_i - z_j)^{-2} K_{ij} -\nu^2
\sum_{i\neq j} (z_i - z_j)^{-2} -\sum_i z_i^2]\,.
\eqno{(26)}
$$

Then one observes that since the Hamiltonian (23) is $S_N$ invariant
$$
[K_{ij} , H ] =0
\eqno{(27)}
$$
it can be restricted to one of the subspaces $L_\pm $
$$
K_{ij} L_\pm = \pm L_\pm
\eqno{(28)}
$$
the restrictions of which we denote as $H^\pm$\,,
$$
H^\pm=
-\half [\Delta
+\nu \sum_{i\neq j} [(z_i -z_j)^{-1} (\dzi -\dzj) T_{ij}
- (z_i -z_j)^{-2} (T_{ij}\mp 1)] -\sum_i z_i^2]\,.
\eqno{(29)}
$$
{}From (26) it finally follows that
$$
g H^\pm g^{-1} = H^\pm_{Cal}\,
\eqno{(30)}
$$
where
$$
H^\pm_{Cal}=
-\half [\Delta
- (\nu^2 \mp \nu)\sum_{i\neq j}(z_i -z_j)^{-2} -\sum_i z_i^2]\, I\,\,.
\eqno{(31)}
$$
and $I$ is the unit matrix in the representation $T$ of $S_N$.

We see that solutions of the Knizhnik-Zamolodchikov equation (24)
interpolate between the eigenfunctions of $H$ (23) spanned by those
linear combinations of the vectors (17) which belong to the subspaces
$L_\pm$ (28) and solutions of the standard Calogero problem described by
the Hamiltonian (31). It is worth mentioning that such a construction
can lead to eigenfunctions of the Calogero Hamiltonian which are not
necessarily normalizable and as such can give rise to some singular
solutions of the generalized harmonic equation by Calogero
$$
[\Delta + \nu \sum_{i\neq j}(z_i -z_j)^{-1} (\dzi - \dzj )] f(z) = 0
\eqno{(32)}
$$
beyond the class of symmetric polynomials
which can be constructed through the Fock-type construction (17) applied to the
Calogero Hamiltonian (31) itself \cite{BHV}.

The above construction can be generalized by tensoring by
some other representation $P$ of $S_N$ ,
$$
H\rightarrow H_P = H \otimes I_P
\eqno{(33)}
$$
where $I_P$ is the unit operator in the representation space $V_P$ of $P$
where the elementary permutations $P_{ij}$ act, and replacing the condition
(28)
by
$$
K_{ij} L_P =P_{ij}  L_P
\eqno{(34)}
$$
where $L_P$ is a subspace of $F\otimes V_P$ where $F$ denotes the Fock space
of the wavefunctions of the original Hamiltonian. It is important that
the matrices $P_{ij}$ commute with the $H_P$ (33) thus ensuring consistency of
its reduction to the subspace (34). The formulae analogous to (30), (31) then
read
$$
g H^P g^{-1} = H^P_{Cal}
\eqno{(35)}
$$
with
$$
H^P_{Cal}=
-\half I\otimes [\Delta
- (\nu^2 - \nu P_{ij})(z_i -z_j)^{-2} -\sum_i z_i^2]
\eqno{(36)}
$$
The Hamiltonian (36) corresponds to the matrix versions of the Calogero model
considered e.g. in \cite{DKV}.

In conclusion let us emphasize that the fact that Knizhnik-Zamolodchikov
equation is related to the Calogero model is rather well known (see e.g.
\cite{Ve1} and references therein). However, the relationship established there
seems to be rather different from the one discussed in this paper. An
 interesting question that remains to be investigated is whether such a
construction can be used in the opposite direction, i.e. to generate the
solutions of the Knizhnik-Zamolodchikov equation as operators $g$ intertwining
between the wave functions of the different versions of the Calogero
Hamiltonians. The fact that our scheme applies to the case of the Calogero
model
with harmonic interaction makes it even more attractive since for this case
there exists the explicit formulae for the wave functions in terms of the Fock
vectors (17).

We believe that the representations of the $S_N$ extended
Heisenberg algebra considered in this paper may have a wider area of
applicability beyond the specific examples considered above.

\vfill

\end{document}